\documentclass[twocolumn,trackchanges]{aastex631}

\usepackage{CJK}
\usepackage{threeparttable}
\usepackage{amsmath}
\usepackage{subfigure}
\usepackage{booktabs}
\usepackage{appendix}
\usepackage{xcolor}
\usepackage{hyperref}
\usepackage{xcolor}
\usepackage{soul}

\graphicspath{{./}{figures/}}
\begin{document}
\begin{CJK*}{UTF8}{gbsn}

\title{EP241021a: A catastrophic collapse/merger of compact star binary leading to the formation of a remnant millisecond magnetar?}

\author[0000-0001-8744-3813]{Guang-Lei Wu(吴光磊)}
\author[0000-0002-1067-1911]{Yun-Wei Yu(俞云伟)}
\author[0000-0002-8708-0597]{Liang-Duan Liu(刘良端)}
\affiliation{Institute of Astrophysics, Central China Normal University, Wuhan 430079, China; \url{yuyw@ccnu.edu.cn}}
\affiliation{Education Research and Application Center, National Astronomical Data Center, Wuhan 430079, China}
%\affiliation{Key Laboratory of Quark and Lepton Physics (Central China Normal University), Ministry of Education, Wuhan 430079, China}
\author[0000-0002-7835-8585]{Zi-Gao Dai(戴子高)}
\affiliation{Deep Space Exploration Laboratory/Department of Astronomy, University of Science and Technology of China, Hefei 230026, People's Republic of China}
\affiliation{School of Astronomy and Space Science, University of Science and Technology of China, Hefei 230026, People's Republic of China}
%\affiliation{School of Astronomy and Space Science, Nanjing University, Nanjing 210023, People's Republic of China}
\author[0000-0003-3440-1526]{Wei-Hua Lei(雷卫华)}
\affiliation{Department of Astronomy, School of Physics, Huazhong University of Science and Technology, Wuhan 430074, People's Republic of China}
\author[0000-0002-6299-1263]{Xue-Feng Wu(吴雪峰)}
\affiliation{Purple Mountain Observatory, Chinese Academy of Sciences, Nanjing 210023, China}
\author[0000-0003-3257-9435]{Dong Xu (徐栋) }
\affiliation{National Astronomical Observatories, Chinese Academy of Sciences, Beijing 100101, China}
\affiliation{Altay Astronomical Observatory, Altay, Xinjiang 836500, China}
\author[0000-0002-9725-2524]{Bing Zhang(张冰)}
\affiliation{Nevada Center for Astrophysics, University of Nevada Las Vegas, NV 89154, USA}
\affiliation{Department of Physics and Astronomy, University of Nevada Las Vegas, NV 89154, USA}
\author[0000-0002-9195-4904]{Jin-Ping Zhu(朱锦平)}
\affiliation{School of Physics and Astronomy, Monash University, Clayton Victoria 3800, Australia}
\affiliation{OzGrav: The ARC Centre of Excellence for Gravitational Wave Discovery, Clayton Victoria 3800, Australia}
\author[0000-0002-5400-3261]{Yuan-Chuan Zou(邹远川)}
\affiliation{Department of Astronomy, School of Physics, Huazhong University of Science and Technology, Wuhan 430074, People's Republic of China}

%\author{}
%\affiliation{}

\begin{abstract}
Observations of fast X-ray transients (FXRTs) with the {\em Einstein Probe} have successfully led to the discovery of some unusual extragalactic optical transients. EP241021a is a newly discovered FXRT that was featured by a significant bump around ten days in both optical and X-ray bands. This timescale and the exceptionally high peak bolometric luminosity up to $\sim \rm 10^{44}erg~s^{-1}$ of the optical bump make it somewhat similar to fast blue optical transients, but still distinctive from them by its relatively red color. We then suggest that the multi-wavelength bump of EP241021a could represent an explosion-type transient, while the underlying power-law decaying component of the optical and X-ray emission as well as the total radio emission are produced by a moderately relativistic jet. 
 By fitting the observed multi-wavelength light curves, it is found that the explosion ejecta that produce the thermal optical emission can have a mass of $\sim0.03~M_{\odot}$, an expanding velocity of $\sim0.25~c$, and an optical opacity of $\sim12~\rm cm^2g^{-1}$, which was continuously powered by a rapidly rotating and highly magnetized neutron star (NS; i.e., a magnetar). In addition to heating the explosion ejecta, the magnetar also provided the dominant contribution to the observed X-ray rebrightening through the non-thermal emission of its wind. These properties suggest that the explosion may result from a catastrophic collapse/merger of a compact star system, which led to the formation of a millisecond magnetar, and the possible progenitor could be an accreting white dwarf (WD) or a binary consisting of double WDs, double NSs, or a WD and an NS.
\end{abstract}

\section{Introduction} \label{sec:intro}
The {\em Einstein Probe} (EP; also known as Tian-Guan in China) satellite is a mission designed to detect transient phenomena in the soft X-ray sky. Two telescopes onboard include the Wide-field X-ray Telescope (WXT), monitoring the sky with its large 3600 deg$^2$ field of view, and the Follow-up X-ray Telescope (FXT), which can quickly characterize and precisely localize
the transients triggered onboard or commanded uplink \citep{Yuan2015,Yuan2022,Yuan2025}. Since its launch on 9 January 2024, EP has successfully captured more than 130 new fast X-ray transients (FXRTs), around 20 of which were found to be associated with a gamma-ray burst (GRB), indicating a potential link between the FXRTs and GRBs. However, for many FXRTs, actually no GRB counterpart is detected, even though they are located in the field of view of gamma-ray telescopes that are sensitive enough to observe a typical GRB \citep{Sun2024,Busmann2025,Ravasio2025,Zhang2025}. Even in the association case such as EP240315a/GRB 240315C, the durations and temporal behaviors of the associated FXRT and GRB could still be very different from each other
\citep{Liu2025,Levan2024}. 
This indicates that the FXRT emission may not simply be considered to be the extension of GRB emission into the low-luminosity range or into the low-energy band.

It is also striking that some FXRTs can further lead to the discovery of an unusual optical transient, e.g., as first reported, EP240414a was found to be followed by an unusual broad-lined Type Ic supernova (SN Ic-BL): SN 2024gsa \citep{Srivastav2025,vanDalen2025,Sun2024}. This situation is similar to that appearing in the XRF 060218/SN 2006aj association event. According to its redshift of $z=0.401$, the luminosity of SN 2024gsa can be inferred to be as high as $\sim2\times10^{43}\rm erg~s^{-1}$, which requires a mass of $0.74^{+0.05}_{-0.04} M_{\odot}$ of $^{56}$Ni if it is powered by radioactivity as usual. This mass of $^{56}$Ni lies at the upper edge of the range of $M_{\rm Ni}$ of GRB-associated SNe \cite[i.e., $\sim0.2-0.6M_{\odot}$\footnote{The measurements of nickle masses for GRB-associated SNe are usually degenerate with the modeling of the simultaneous jet afterglow emission, which can lead to a systematic uncertainty of the nickle masses.};][]{Cano2013, Prentice2016, Cano2017, Lv2018}, indicating that SN 2025gsa requires an unusually large amount of $^{56}$Ni synthesis. It should be noted that the range of $\sim0.2-0.6M_{\odot}$ is already several times higher than that of normal core-collapse SNe which is usually no more than $\sim0.2M_{\odot}$ \citep{Woosley2006, Hjorth2012}.
In addition, before SN 2024gsa, another significant optical light curve bump was found to appear two days after the FXRT, which rises very quickly to a peak absolute magnitude of $-21$ AB mag within a few days. The origin of this early bump is debated \citep{Sun2024,Hamidani2025}. Very recently, such a double-bump optical light curve was also discovered following FXRT EP250108a \citep{Rastinejad2025, Eyles2025, Srinivasaragavan2025, Li2025}, where a more luminous SN (i.e., SN 2025kg) was accompanied by weaker X-ray emission.

The SNe following FXRTs EP240414a and EP250108a are both too luminous to be accounted for solely by the traditional radioactive power, which makes them potentially connected to GRBs as well as superluminous supernovae (SLSNe). Therefore, a central engine could be expected to play a crucial role in driving these FXRTs, since such an engine (in the particular sense, a rapidly rotating magnetar) was usually used to explain the phenomena of GRBs and SLSNe \cite[e.g.,][]{Dai1998, Zhang2001, Yu2007, Kasen2010, Metzger2015, Yu2017, Margalit2018, Fiore2025}. Furthermore, the big bump on a timescale of several days, which is before the SN peak, could provide more detailed clues to the explosion processes and environments of the FXRTs. First,  the bump could be the result of the cooling of SN ejecta and dense circum-stellar material that were previously heated by the SN shock \citep{Sun2024,vanDalen2025,Srinivasaragavan2025}, in particular, when the SN shock can be energized by the powerful wind of a magnetar \citep{Zhang2022}. Nevertheless, it was also suggested that the bump could have a non-thermal origin, e.g., due to a shock driven by a jet and refreshed by the jet tail or due to a shock driven by an off-axis cocoon \citep{Hamidani2025,Zheng2025}. In the former scenario, it is assumed that the velocity stratification of the jet enables slower shells to catch up with the decelerating shock front, thereby re-energizing the front and inducing a delayed brightening \citep{Rees1998,Sari2000,Srivastav2025}.

Very recently, a significant light curve bump has been further discovered from EP241021a, which appeared ten days after the FXRT emission \citep{Busmann2025,Gianfagna2025,Shu2025,Yadav2025}. 
The timescale of this bump is just shorter than the typical timescale of SNe of about several tens of days, but relatively longer than the cooling time of SN ejecta, which is on the order of a few days. Meanwhile, the absolute peak magnitude of the optical bump can be set at $M_{\mathrm{r}}\approx-22$ mag, according to its relatively high redshift $z=0.748$ \citep{Pugliese2024, Perez-Fournon2024, Zheng2024}. This luminosity and timescale of the optical bump make it resemble some mysterious fast blue optical transients \citep[FBOTs;][]{Drout2014} such as AT 2018cow, although the optical color of EP 241021a is redder. It is worth mentioning that AT 2018cow was prominent in the FBOT population due to its abundant X-ray emission \citep{Perley2019, Rivera2018, Margutti2019}, while the optical bump of EP241021a is also accompanied by an X-ray rebrightening that is completely different from EP240414a and EP250108a. The evolution behavior of the X-ray emission of AT 2018cow \citep{Margutti2019} and the possible quasi-periodic oscillation in it \citep{Pasham2022} further shows a strong potential connection between this X-ray emission and the spin down of a millisecond magnetar. Therefore, in this paper, we focus on the possibility that the multi-wavelength bump of EP241021a was contributed by an explosion-type transient, in view that some other types of model have been previously proposed including a refreshed shock \citep{Busmann2025}, a structured jet as well as its cocoon \citep{Gianfagna2025}, and an intermittent energized shock due to repeating partial tidal disruption event involving an intermediate-mass black hole \citep{Shu2025}.

\begin{figure}
    \centering
    \includegraphics[width=1\linewidth]{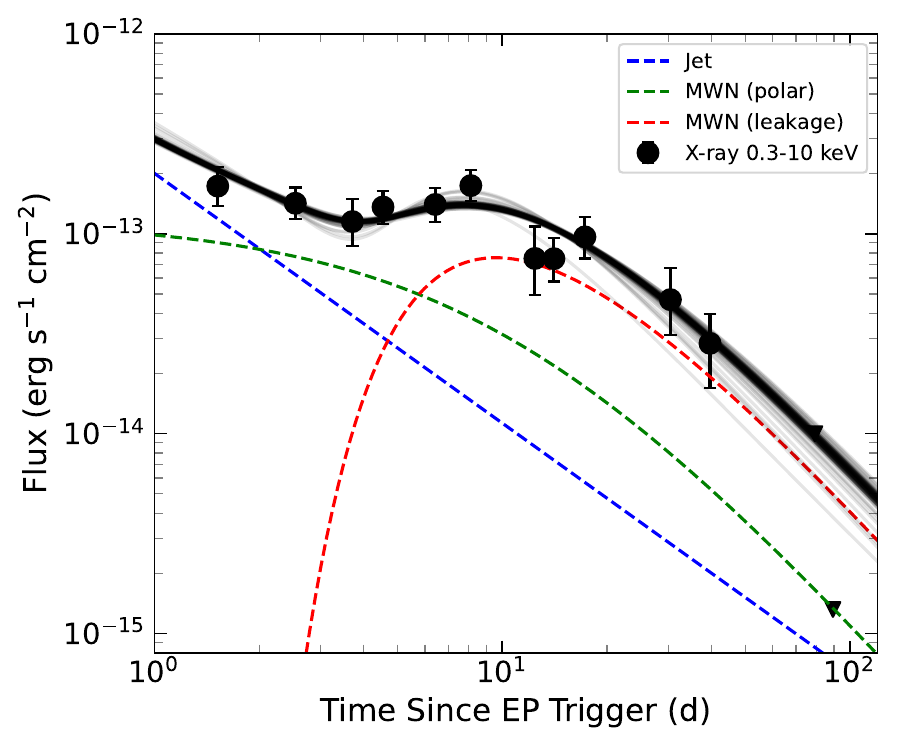}
    \caption{Observed X-ray light curve of EP241021a (solid circles) and the model fittings (lines). The data are taken from \cite{Shu2025}.}
    \label{fig:xlc}
\end{figure}

\section{Observation of EP241021a}\label{obs}

EP241021a was initially detected by the WXT onboard EP at 05:07:56 UTC with a time-averaged X-ray flux of $3.31^{+1.26}_{-0.86}\times 10^{-10} \mathrm{erg~cm^{-2}~s^{-1}} $ in the 0.5--4.0 keV photon energy range. The transient has a duration of $\sim90~\mathrm{s}$ and a relatively hard X-ray spectrum with a photon index of $1.8^{+0.57}_{-0.54}$ \citep{Shu2025}. No high-energy gamma-ray counterpart to EP241021a was detected by Fermi-GBM \citep{Burns2024} or Konus-Wind \citep{Svinkin2024}, despite both targeted and blind searches around the trigger time. Specifically, Fermi-GBM places a $3\sigma$ upper limit on the fluence in the $30-400~\mathrm{keV}$ band of $9.8\times10^{-7}\mathrm{erg~cm^{-2}}$ \citep{Gianfagna2025} and Konus-Wind reports a $90\%$ confidence upper limit on the $20-1,500~\mathrm{keV}$ peak flux of $2.5\times10^{-7}\mathrm{erg~cm^{-2}s^{-1}}$ in a timescale of $2.994~\mathrm{s}$ \citep{Svinkin2024}. These gamma-ray non-detections constrain the spectral peak energy of EP 241021a to $E_{\mathrm{p}} \lesssim 100~\mathrm{keV}$ and the isotropic-equivalent energy to $E_{\mathrm{iso}} \lesssim 2 \times 10^{51}~\mathrm{erg}$, where a Band function extrapolated from the WXT detection is assumed.
After 36.5 hours of the initial detection, EP241021a was observed by the FXT on board EP, with a total of 12 observations conducted \citep{Shu2025}. Follow-up observations were also performed with XMM-Newton, yielding an upper limit at $T_0+89$ days \citep{Shu2025}. 
As shown in Figure \ref{fig:xlc}, an X-ray light curve was revealed, which contains a hint of an initial decay, a plateau with a possible slight rise between $T_0+3$ and $T_0+10$ days, and a second decay roughly following a behavior of $\sim t^{-2}$ until an abrupt decline around $T_0+80$ days.

\begin{figure}
    \centering
    \includegraphics[width=1\linewidth]{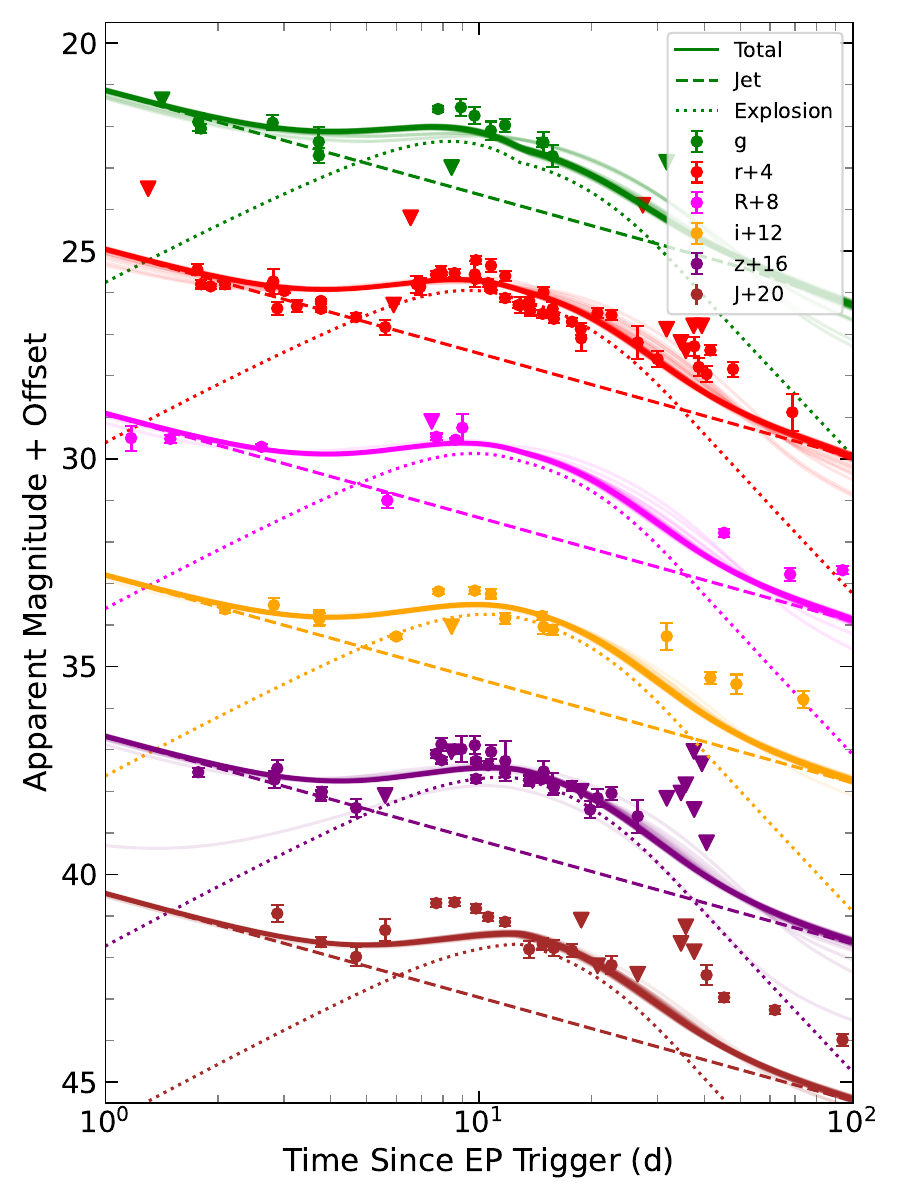}
    \caption{Observed optical light curves of EP241021a (solid circles) and the model fittings (lines). The data are taken from \cite{Busmann2025} and \cite{Shu2025} but with an extra correction of $E(B-V)=0.045~\mathrm{mag}$ and  $R_{{V}}=3.1$ corresponding to the Galactic extinction \citep{Schlafly2011}.}
    \label{fig:optlc}
\end{figure}

Optical photometry of EP241021a was first reported by Nordic Optical Telescope at R.A. = 01:55:23.41, Dec. = +05:56:18.01 with an uncertainty of $0.5''$ \citep{Fu2024} and was confirmed by other earlier observations. The following observations were performed using multiple instruments \citep[e.g., Thai Robotic Telescope, Katzman Automatic Imaging Telescope, Liverpool Telescope, ALT-50A\&-100C, HMT-0.5m Telescope, Gran Telescopio Canarias, Fraunhofer Telescope Wendelstein, Hobby-Eberly Telescope, Very Large Telescope, The Swift Ultra-Violet Optical Telescope, Calar Alto Telescope, Large Binocular Telescope; ][]{Busmann2025,Gianfagna2025,Shu2025}. 
As presented in Figure \ref{fig:optlc}, the initial decay of the optical light curves could basically trace the early X-ray emission, while a significant bump appears in all bands coinciding with the X-ray plateau/rebrightening. Besides the first bump around $T_0+10$ days, a second bump seems to exist at $T_0+50$ days (see the argument in \citealt{Shu2025}) although the signal is still ambiguous.
The optical spectroscopy of EP241021a was conducted by Gran Telescopio Canarias \citep{Perez2024}, Keck I 10 m telescope \citep{Zheng2024}, Very Large Telescope \citep{Pugliese2024}, and Hobby-Eberly Telescope \citep{Busmann2025}. The detected narrow emission lines of [O II] and [O III] are likely from the host galaxy and give the redshift of $z=0.748$, while no SN-like features are identified in these spectra. 

\begin{figure}
    \centering
    \includegraphics[width=1\linewidth]{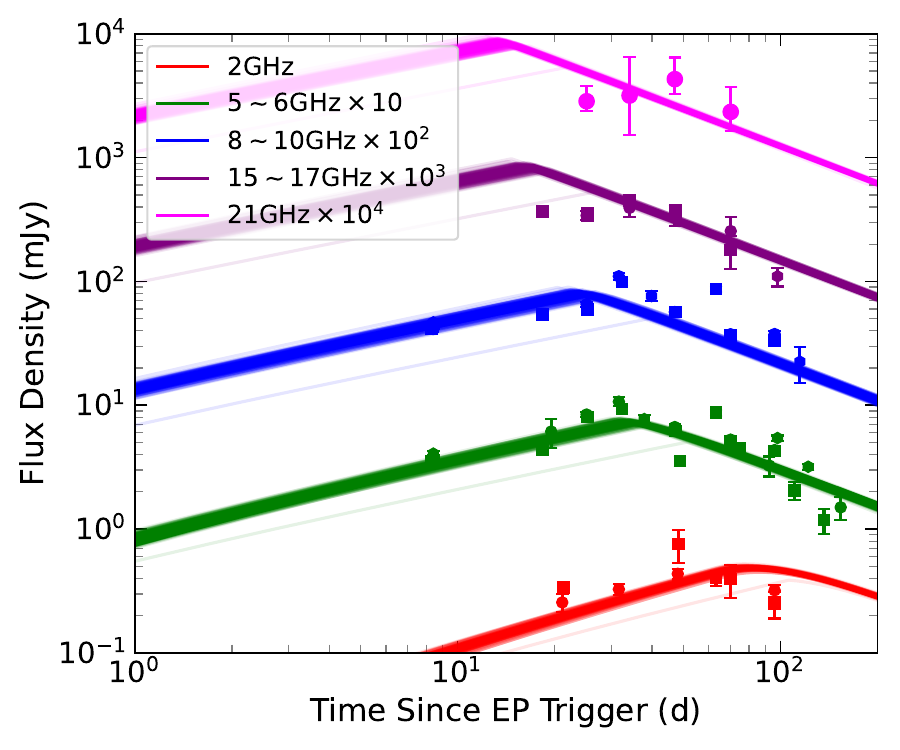}
    \caption{Radio emission of EP241021a and the fitting with a jet shock emission. The data are taken from  \cite{Yadav2025}, \cite{Gianfagna2025}, and \cite{Shu2025}, which are represented by squares, circles, and hexagrams, respectively. Despite sharing the same color labeling, data from different references may exhibit subtle frequency differences. Accordingly, theoretical curves are plotted over the corresponding frequency ranges, with model parameters sampled from their posterior distributions.}
    \label{fig:rlc}
\end{figure}

Finally, radio observations were carried out with ATCA \citep{Gianfagna2025,Yadav2025}, e-MERLIN \citep{Gianfagna2025,Yadav2025},
uGMRT, ALMA \citep{Gianfagna2025},
VLA, MeerKAT, and VLBA \citep{Shu2025}, simultaneously. 
The temporal evolution of the radio emission can be roughly described by a power-law rise and decline with a peak shifting toward lower frequencies over time, as displayed in Figure \ref{fig:rlc}.

\section{The model}\label{model}

\subsection{An explosion transient emission}
Prior to detailed modeling of the multi-wavelength light curves of EP 241021a, we first conduct an order-of-magnitude analysis of the optical bump appearing at 10 days. First of all, the peak bolometric luminosity of the bump can be calculated to $L_{\rm p}\sim 10^{44}\mathrm{erg~s^{-1}}$ with a black body assumption which will be checked in Section \ref{sec: fitting}. The blackbody temperature can be estimated to be within the range of $T = (1+z)h\nu/(2.82k_{\mathrm{B}})=(0.7\sim 1.7)\times 10^{4} ~\mathrm K $, where the spectral peak frequency is considered to be located in the observational range of $(4\sim10 )\times 10^{14}~\mathrm{Hz}$, $k_{\mathrm{B}}$ and $h$ are the Boltzmann constant and the Planck constant, respectively. Then, the radius of the emitting photosphere can be estimated as
\begin{equation}
    \begin{aligned}
        &R_{\mathrm{ph}}=\left(\frac{L_{\rm p}}{4\pi \sigma_{\mathrm{SB}}T^4}\right)^{1/2}\\
        &\simeq3.8\times 10^{15}\left(\frac{L_{\rm p}}{10^{44}~\mathrm{erg~s^{-1}}}\right)^{1/2}\left(\frac{T}{10^{4}~\mathrm{K}}\right)^{-2}\mathrm{cm}, 
    \end{aligned}  
\end{equation}
where $\sigma_\mathrm{SB}$ is the Stefan-Boltzmann constant. In the stellar explosion-type scenario, the photosphere can expand along with the expansion of the explosion ejecta during the rising phase of emission. In this case, because the ejecta is very optically thick, the outer radius of the ejecta can be close to but slightly larger than the photospheric radius, i.e., $R_{\rm ej}\gtrsim R_{\rm ph}$. With the temperature range considered above, we can further constrain the expansion velocity of the ejecta to be $v_{\mathrm{ej}}>R_{\rm ph}/t_{\rm p}\simeq(0.08\sim0.5)~c$, where  $t_{\rm p}\simeq10/(1+z)\simeq 6$ days is the peak time of the optical bump. In view of the fact that the energy of prompt X-ray emission is about $10^{50}\mathrm{erg}$, we tentatively take the kinetic energy of the ejecta to be $E_{\mathrm{k}}\sim10^{51}\mathrm{erg}$ as a reference value. The more accurate value of this energy depends on the determination of the efficiency of the prompt emission. In any case, the value of $\sim10^{51}\mathrm{erg}$ is still very typical for both SN explosions and compact object mergers. So, the mass of the ejecta can be estimated to $M_{\mathrm{ej}}=(0.005\sim0.15)~M_{\mathrm{\odot}}$. Then, according to the following expression for the photon diffusion timescale of an expanding ejecta: 
 \begin{equation}
    \begin{aligned}
       & t_{\mathrm{diff}}\simeq\left(\frac{3\kappa_{\rm opt} M_{\mathrm{ej}}}{4\pi c v_{\mathrm{ej}}}\right)^{1/2}\\
        &\simeq 6.5\left(\frac{\kappa_{\rm opt}}{10~\mathrm{{cm}^2g^{-1}}}\right)^{1/2}\left(\frac{M_{\mathrm{ej}}}{0.02~M_{\odot}}\right)^{1/2}\left(\frac{v_{\mathrm{ej}}}{0.25~c}\right)^{-1/2}~\mathrm{day},
    \end{aligned}
 \end{equation}
we find that a relatively high opacity $(\kappa_{\rm opt}\sim 10~\mathrm{{cm}^2g^{-1}})$ would be invoked to match the observed peak time.

The above analyses suggest that the optical bump after EP241021a is unlikely to be produced by a normal SN explosion, although the possibility of it being an ultra-stripped SN could still exist if the ejecta mass can reach the upper bound of the estimates \citep{Tauris2013a, Suwa2015, Tauris2015, Muller2018}. In any case, by comparison, the properties of the explosion ejecta of EP241021a seem to be more consistent with those predicted for the material ejected from the collapse of super-Chandrasekhar white dwarfs \citep[WDs;][]{Dessart2006, Batziou2024, Longo2023, Cheong2025, Kuroda2025},  double neutron star (DNS) mergers \citep[e.g.,][]{ Hotokezaka2013, Sekiguchi2016, Radice2016, Radice2018,  Radice2022, Dietrich2017a, Dietrich2017b, Shibata2019,Combi2023}, and NS-WD mergers \citep{Metzger2012,Fernandez2013,Margalit2016,Fernandez2019,Zenati2019,Zenati2020}. The possible neutron-rich property of this material would be very conducive to the formation of r-process elements, particularly in the DNS merger case. Among these elements, lanthanides could lead to an unusually high opacity such as $(10\sim100)~\rm cm^2g^{-1}$ \citep{Kasen2013,Tanaka2013, Tanaka2020,Even2020}, which gives a natural explanation for the required opacity of the optical bump of EP241021a. The remaining problem is that the low mass of the ejecta severely limits the total amount of radioactive elements inside, making it impossible to provide the total energy of $\sim10^{50}$ erg of the optical bump. This situation also presents in the FBOT phenomenon and then a more powerful energy source would be involved. %, which is most likely offered by the remnant compact object. 
Specifically, in view of its appropriate amount and timescale of energy release, a millisecond magnetar has been widely suggested to be the central engine of many super-luminous transients \citep[e.g.,][]{Kasen2010, Yu2013, Metzger2014, Yu2015, Gao2015,Yu2019}. And such a magnetar could indeed also be formed from DNS/NS-WD mergers \citep{Dai2006,Wang2024,Sun2025} and collapses of super-Chandrasekhar WDs \citep{Margalit2019,Zhu2021,Yang2022,Longo2023,Kuroda2025}, where a super-Chandrasekhar WD could evolve from an accreting WD in binary with a non-degenerate companion or from a double WD merger event.

Therefore, first of all, we express the energy supply for the optical transient by according to the spin-down of a magnetar, which gives 
\begin{eqnarray}
L_{\mathrm{spl}}(t)&=& \xi L_{\rm sd}(t) \nonumber\\
&=&\xi L_{\mathrm{sd,i}}\left(1+\frac{t}{t_{\mathrm{sd}}}\right)^{-2},
\end{eqnarray}
where the initial value of the spin-down luminosity and the spin-down timescale are expressed as usual as $L_{\mathrm{sd,i}}= 10^{47}B_{\mathrm{p,14}}^2P_{\mathrm{i,-3}}^{-4}~ \mathrm{erg~s^{-1}}$ and $t_{\mathrm{sd}}= 2\times10^5 ~(1+z)B_{\mathrm{p,14}}^{-2} P_{\mathrm{i,-3}}^2~\mathrm{s}$ with $B_{\mathrm{p}}$ and $P_{\mathrm{i}}$ being the magnetic field strength and initial spin period of the magnetar, respectively. The convention $Q_{x}=Q/10^{x}$ is adopted in cgs units. In our calculations, we take $L_{\mathrm{sd,i}}$ and $t_{\mathrm{sd}}$ as free parameters directly. The coefficient $\xi = 1 - \exp(-\tau_{\mathrm{x}})$ is introduced to represent the fraction of the spin-down energy absorbed by the explosion ejecta, where $\tau_{\mathrm{x}} = 3\kappa_{\mathrm{x}} M_{\mathrm{ej}} / (4\pi v_{\mathrm{ej}}^2 t^2)$ is the optical depth for X-rays with $\kappa_{\mathrm{x}}$ being the effective X-ray opacity. 
It is considered that the spin-down energy released from the magnetar is initially in the form of a Poynting flux and subsequently carried out by an ultra-relativistic electron-positron wind of a Lorentz factor $\sim 10^{3-7}$ \citep{Kennel1984,Kirk2009,Bucciantini2011,Amato2020}. This energy conversion occurs mainly through magnetic reconnection processes, which can simultaneously produce X-ray emission \citep{Giannios2005,Beniamini2014,Sironi2014,Kagan2015}. When the magnetar wind catches up and collides with the previous ejecta, a reverse shock can be formed in the wind to convert its kinetic energy into heat. The expansion velocity of the reverse-shocked wind (i.e., the magnetar wind nebula; MWN) is continuous with that of the ejecta at the contact discontinuity. The hot electron-positron plasma in the MWN can release its internal energy through the synchrotron and inverse-Compton processes of relativistic leptons. The emitted photons would be mainly in X-rays too \citep{Metzger2014, Yu2019}, at least, at early times. Therefore, the X-ray opacity is involved in the absorption coefficient.  

Driven by the spin-down energy, the explosion ejecta can be heated effectively to produce a bright thermal emission, the bolometric luminosity of which can be approximately calculated by the following formulae \citep{Arnett1982}:  
\begin{equation}
    L_{\mathrm{th}}= e^{-(t/t_{\mathrm{diff}})^2}\times\int^{t}_{0}2 L_{\mathrm{spl}}\left(t'\right) \frac{t'}{t^2_{\mathrm{diff}}}e^{(t'/t_{\mathrm{tiff}})^2}dt,\label{eq:Lth}
\end{equation}
where $t_{\mathrm{diff}}$ is the photon diffusion timescale of the ejecta. The black-body temperature of this thermal emission can further be given by 
\begin{equation}
T(t) = \max \left[ \left( \frac{L_{\mathrm{th}}}{4\pi \sigma_{\mathrm{SB}} R_{\mathrm{ph}}^2} \right)^{1/4}, T_{\mathrm{floor}} \right],
\end{equation}
where the photospheric radius is determined by $R_{\mathrm{ph}}\approx v_{\mathrm{ej}} t$ with a constant expansion velocity $v_{\rm ej}$ and $T_{\mathrm{floor}}$ is taken as a free parameter. Then, the monochromatic luminosity of the ejecta emission at an observed frequency $\nu$ can be given by 
\begin{equation}
    L_{\rm th, \nu} = (1+z)\frac{8\pi^2R_{\mathrm{ph}}^2}{c^2}\frac{ h [(1+z)\nu]^3}{e^{(1+z)h\nu/k_{B}T}-1}.
\end{equation}

Here, it should be pointed out that, as the ejecta expands, the X-ray emission generated from the MWN can increasingly penetrate the ejecta to be detected. Thus, a rebrightening of the X-ray emission accompanying the optical bump can be naturally understood.
Such a feature makes EP241021a intrinsically different from the other analogous transient events such as EP240414a and EP250108a. In other words, the modeling of the X-ray emission of EP241021a would play a crucial role in identifying its origin. Additionally, if the total spin-down energy can finally exceed the initial kinetic energy of the ejecta, then it needs to be taken into account that the ejecta velocity $v_{\rm ej}$ can increase with time.

\subsection{The external shock of a jet}
The existence of the bright radio emission and the underlying power-law decaying component of the radio and X-ray emission strongly indicate that a powerful jet could be driven by the WD collapses or DNS/NS-WD mergers, which gives rise to multi-wavelength non-thermal emission through an external shock by interacting with the surrounding medium. On the one hand, the relativistic degree of this jet could not be as high as those of normal GRBs, at least on the viewing direction, because no GRB emission was detected accompanying the FXRT emission. On the other hand, the jet could in principle have a complicated angular structure (e.g., as conceived in \citealt{Gianfagna2025}), which would be beneficial for explaining the detailed features appearing in the observed radio light curves. Nevertheless, in this paper, we take into account a top-hat jet simply, which would be found to be good enough to describe the general evolution behavior of the underlying non-thermal emission in all of the radio, optical, and X-ray bands.

The interaction of a relativistic jet with the surrounding medium would initially drive a reverse shock propagating into the jet and a forward shock propagating into the medium. After the reverse shock crosses the whole jet during the first few hundreds of seconds, the forward shock (i.e., a blast wave) would enter into the Blandford-McKee self-similar evolution phase, during which the bulk Lorentz factor of the shock can be written as \citep{Blandford1976,Huang1999}: 
\begin{equation}
\Gamma_{\mathrm{j}}(t) = \left[ \frac{17 E_{\mathrm{j}}(1+z)^3}{1024 \pi n_0 m_{\mathrm{p}} c^5 t^3} \right]^{1/8},
\end{equation}
where $E_{\mathrm{j}}$ is the isotropically equivalent kinetic energy of the jet, $n_0$ is the particle number density of the interstellar medium,
and $m_{\mathrm{p}}$ is the proton mass. 
The electrons, which are accelerated by the forward shock, would be distributed on its comoving random Lorentz factor as a power law $\gamma^{-p}$ with a minimum value of $\gamma_{\mathrm{m}}=\epsilon_{\mathrm{e}}(\Gamma_{\mathrm{j}}-1)\left( p-2\right)m_{\mathrm{p}}/[\left(p-1\right)m_{\mathrm{e}}]$, where $\epsilon_{\mathrm{e}}$ is the equipartition factor for the electrons. Considering the modification of the electron distribution due to their synchrotron cooling, a cooling Lorentz factor is usually defined as $\gamma_{\mathrm{c}}=(1+z)6\pi m_{\mathrm{e}}c/\sigma_{T}B_{\mathrm{j}}'^2\Gamma_{\mathrm{j}} t$, where $B'_{\mathrm{j}}=\sqrt{32\pi \epsilon_{B}\Gamma_{\mathrm{j}}^2 n_{0}m_{\mathrm{p}}c^2} $ is the strength of the stochastic magnetic field with $\epsilon_{B}$ being its equipartition factor.  

Following \cite{Sari1998}, we can further define two characteristic frequencies as $\nu_{\mathrm{m}}=\Gamma_{\rm j}q_{\mathrm{e}}B'_{\mathrm{j}}\gamma_{\mathrm{m}}^2/[2\pi m_{\mathrm{e}}c(1+z)]$ and $\nu_{\mathrm{c}}=\Gamma_{\rm j}q_{\mathrm{e}}B'_{\mathrm{j}}\gamma_{\mathrm{c}}^2/[2\pi m_{\mathrm{e}}c(1+z)]$ for the synchrotron emission spectrum. To be specific, the spectrum can be expressed as
\begin{equation}
\begin{aligned}
   L_{\mathrm{j},\nu} =&  L_{\mathrm{j,\nu,max}} \frac{(1-e^{-\tau_{\mathrm{ssa}},\nu})}{\tau_{\mathrm{ssa},\nu}} \\   
&\times\begin{cases} 
\left( \frac{\nu}{\nu_\mathrm{l}} \right)^{1/3}, & \nu < \nu_\mathrm{l}; \\[10pt]
\left( \frac{\nu}{\nu_\mathrm{l}} \right)^{-(q-1)/2}, & \nu_\mathrm{l} < \nu < \nu_\mathrm{h}; \\[10pt]
\left( \frac{ \nu_\mathrm{h}}{\nu_\mathrm{l}} \right)^{-(q-1)/2} \left( \frac{\nu}{ \nu_\mathrm{h}} \right)^{-p/2}, &  \nu_\mathrm{h} < \nu,
\end{cases}
\end{aligned}
\end{equation}
where $\nu_{\mathrm{l}}=\mathrm{min}\left[\nu_{\mathrm{m}},\nu_{\mathrm{c}}\right]$ and $\nu_{\mathrm{h}}=\mathrm{max}\left[\nu_{\mathrm{m}},\nu_{\mathrm{c}}\right]$, $q=2$ for $\nu_{\mathrm{m}}>\nu_{\mathrm{c}}$ and $q=p$ for $\nu_{\mathrm{m}}<\nu_{\mathrm{c}}$. The spectral peak luminosity reads $L_{\mathrm{j,\nu,max}} = (1+z)\Gamma_{\rm j}N_{\mathrm{e}}m_{\mathrm{e}}c^2\sigma_{T} B'_{\mathrm{j}}/3q_{\mathrm{e}}$,

where $m_{\mathrm{e}}$, $q_{\mathrm{e}}$ and $\sigma_{\mathrm{T}}$ are the electron mass, charge, and the Thomson cross-section, respectively. The total number of electrons is given by $N_{\mathrm{e}}=4\pi R_{\mathrm{j}}^3 n_0/{3} $ with $ R_{\mathrm{j}}=[17E_{\mathrm{j}}t/4\pi (1+z)m_{\mathrm{p}}n_0c]^{1/4}$. Finally, the synchrotron self-absorption depth can be expressed as
\begin{equation}
\begin{aligned}
  \tau_{\mathrm{ssa},\nu} = &C\frac{q_{\mathrm{e}}N_{\mathrm{e}}}{4\pi B'_{\mathrm{j}}R_{{\mathrm{j}}}^2\gamma_{\mathrm{l}}^{5}}\\
    &\times\begin{cases} 
\left( \frac{\nu}{\nu_\mathrm{l}} \right)^{-5/3}, & \nu < \nu_\mathrm{l}; \\[10pt]
\left( \frac{\nu}{\nu_\mathrm{l}} \right)^{-(q+4)/2}, & \nu_\mathrm{l} < \nu < \nu_\mathrm{h}; \\[10pt]
\left( \frac{ \nu_\mathrm{h}}{\nu_\mathrm{l}} \right)^{-(q+4)/2} \left( \frac{\nu}{ \nu_\mathrm{h}} \right)^{-(q+5)/2}, &  \nu_\mathrm{h} < \nu.
\end{cases}
\end{aligned}
\end{equation}
where $\gamma_{\mathrm{{l}}}= \mathrm{min}[\gamma_{\mathrm{m}},\gamma_{\mathrm{c}}]$  and $C\approx5$ is a factor slightly depending on $p$ \citep{Panaitescu2000}.

\section{Light curve fittings}\label{sec: fitting}
\begin{figure}
    \centering
    \includegraphics[width=1\linewidth]{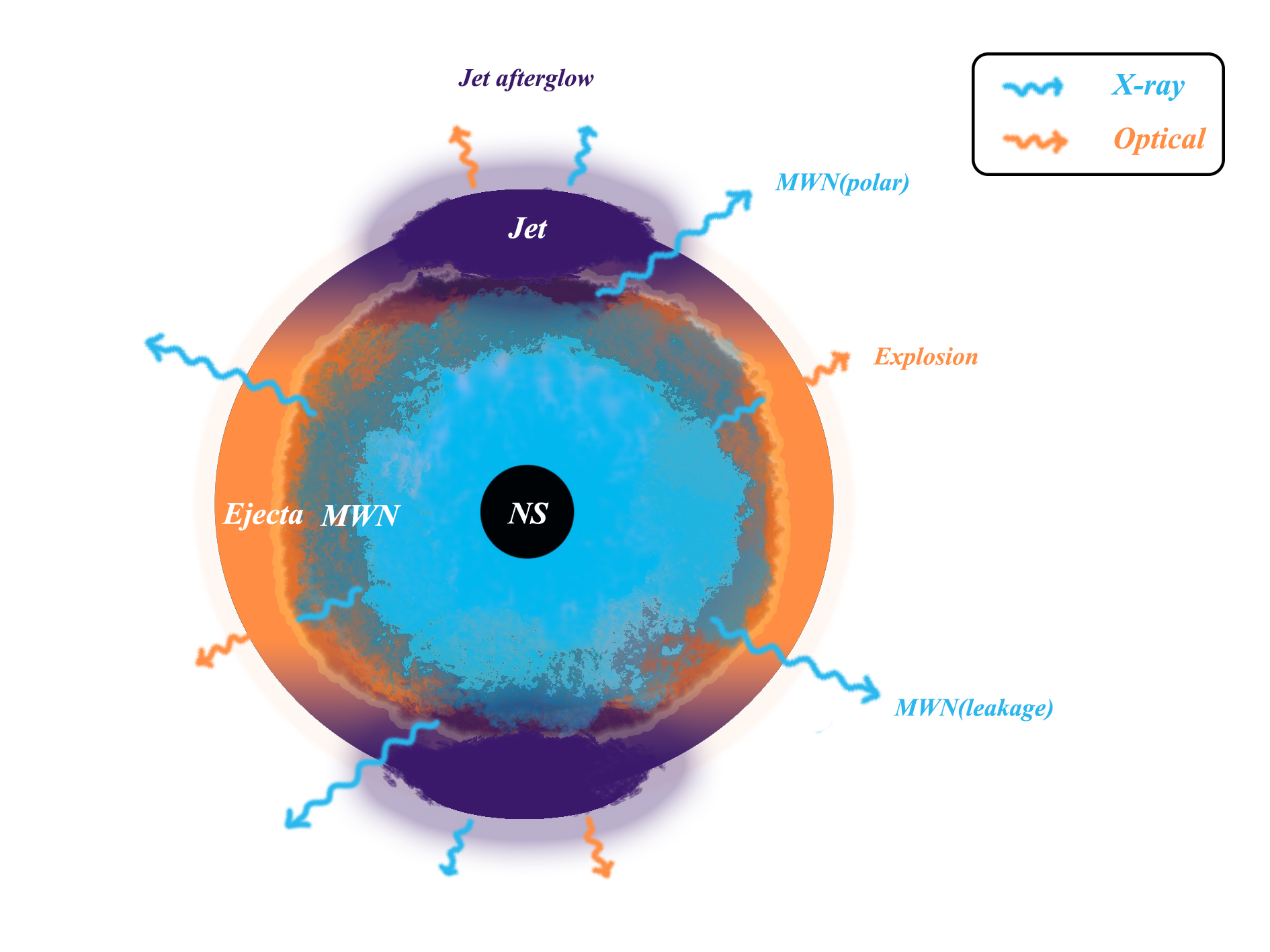}
    \caption{An illustration of the EP241021a explosion.}
    \label{fig:ill}
\end{figure}

In our model, the radio emission after FXRT EP241021a is considered to be purely contributed by the external shock of a jet, while the optical emission has two origins including the non-thermal emission of the jet and the thermal emission of the explosion ejecta. So, we have
\begin{equation}
L_{\rm opt,tot}=L_{\mathrm{th},\nu}+L_{\mathrm{j,\nu}}.\label{eq:Lopt}
\end{equation}
In order to confront with the observational data, the flux density of the optical emission at observed frequency $\nu_{}$ can be given by $ F_{\nu} = L_{\nu}/4\pi D_{L}^2$, where $D_{L}$ is the luminosity distance. The apparent magnitude is given by $M_{\nu}=-2.5~\mathrm{log_{10}}\left(F_{\nu}\right)-48.6$ in the AB system where $F_{\nu} $ is in units of $\mathrm{erg~s^{-1}cm^{-2}Hz^{-1}}$. By comparison, the generation of the X-ray emission could be most complicated. As illustrated in Figure~\ref {fig:ill}, besides the contribution from the jet, the emission leaking from the MWN and the magnetic reconnection zone could gradually become the dominant component of the X-ray emission, which is further direction-dependent. In our calculations, the observer is assumed on the axis of the jet.

On the one hand, the MWN emission itself could be anisotropic, so we introduce a free parameter $\zeta$ to represent the fraction of the spin-down energy released into the jet direction. If $\zeta \rightarrow 1$, then the energy is highly beamed in the jet direction. In this case, an ultra-relativistic jet could be easily launched. Otherwise, the energy is anti-beamed for $\zeta\ll \Omega_{\rm j}/4\pi$, which could be unfavorable to the formation of an ultra-relativistic jet, where $\Omega_{\rm j}$ is the solid opening angle of the jets. On the other hand,  Moreover, the leakage of the MWN emission is further dependent on the anisotropic structure of the jet and ejecta. In the jet direction, the MWN emission is likely to be detectable from the very moment of the beginning, because the jet material is always transparent. On the contrary, in the other directions, the MWN emission can be detected only after the explosion ejecta becomes optical thin for X-rays. Following the above considerations, the total luminosity in X-ray band can be expressed as  
\begin{equation}
L_{\mathrm{X,tot}} =L_{\mathrm{j,X}}+{\zeta\over f_{\rm j}} L_{\mathrm{sd}}+{(1-\zeta)}L_{\mathrm{sd}} e^{-\tau_X},\label{eq:LX}
\end{equation}
where the X-ray luminosity of the MWN is simply assumed to trace the spin-down luminosity of the magnetar without a detailed description of the MWN emission mechanisms. $f_{\rm j}=\Omega_{\rm j}/4\pi$ is introduced because the MWN at the jet direction is probably relativistic and, consequently, its emission could be collimated within such a solid angle rather than isotropic. Here we take $f_{\rm j}\approx 0.15$ as a reference value, roughly corresponding to a half-opening angle around $30^\circ$ of the jet.

\begin{table}[htbp]
     \centering
     \begin{tabular}{cccc}
     \hline
     \hline
        Parameter   & Prior & Allowed Range & Posteriors\\
        \hline
        &External Shock&&\\
        \hline
    $E_{\rm j}/10^{51} \mathrm{erg}$ & Log-flat & [0.01,100] & $2.37_{-0.17}^{+0.18}$ \\
    $n_{0}/ \mathrm{cm^{-3}}$ & Log-flat & [0.001,10] & $0.28_{-0.07}^{+0.10}$ \\
    $p$ & Flat & [2,3] & $2.34_{-0.01}^{+0.01}$ \\
    $\epsilon_{\mathrm{e}}$ & Flat & [0.01,1] & $0.15_{-0.01}^{+0.01}$ \\
    $\epsilon_{B}$ & Log-flat & [0.0001,0.1] & $0.03_{-0.01}^{+0.01}$ \\
        \hline
        &Isotropic Ejecta&&\\
        \hline
    $M_{\mathrm{ej}}/M_{\odot}$ & Log-Flat & [0.001,0.5] & $0.03_{-0.01}^{+0.02}$ \\
    $v_{\mathrm{ej}}/c$ & Flat & [0.01,0.5] & $0.25_{-0.02}^{+0.02}$ \\
    $\kappa/\mathrm{cm^2g^{-1}}$ & Flat & [0.1,50] & $12.58_{-4.52}^{+3.93}$ \\
    $\kappa_{\mathrm{x}}/\mathrm{cm^2g^{-1}}$ & Flat & [0.1,10] & $1.83_{-0.65}^{+0.63}$ \\ 
    $T_{\mathrm{floor}}/10^3 \mathrm{K}$ & Flat & [1,10] & $8.19_{-0.19}^{+0.20}$ \\    
        \hline
        &Magnetar&&\\
        \hline
    $L_{\rm sd,i}/\mathrm{10^{47}~erg~s^{-1}}$ & Log-Flat & [0.001,1000] & $0.01_{-0.002}^{+0.002}$ \\
    $t_{\mathrm {sd}}/10^{5}\rm s$ & Log-Flat & [0.001,1000] & $5.19_{-0.51}^{+0.52}$ \\
    $\zeta$ & Log-Flat & [0.001,1] & $0.04_{-0.005}^{+0.005}$ \\
     \hline
     \end{tabular}
     \caption{The priors and posteriors of model parameters.}
     \label{para_tab}
\end{table}

\begin{figure*}
    \centering
    \includegraphics[width=1\linewidth]{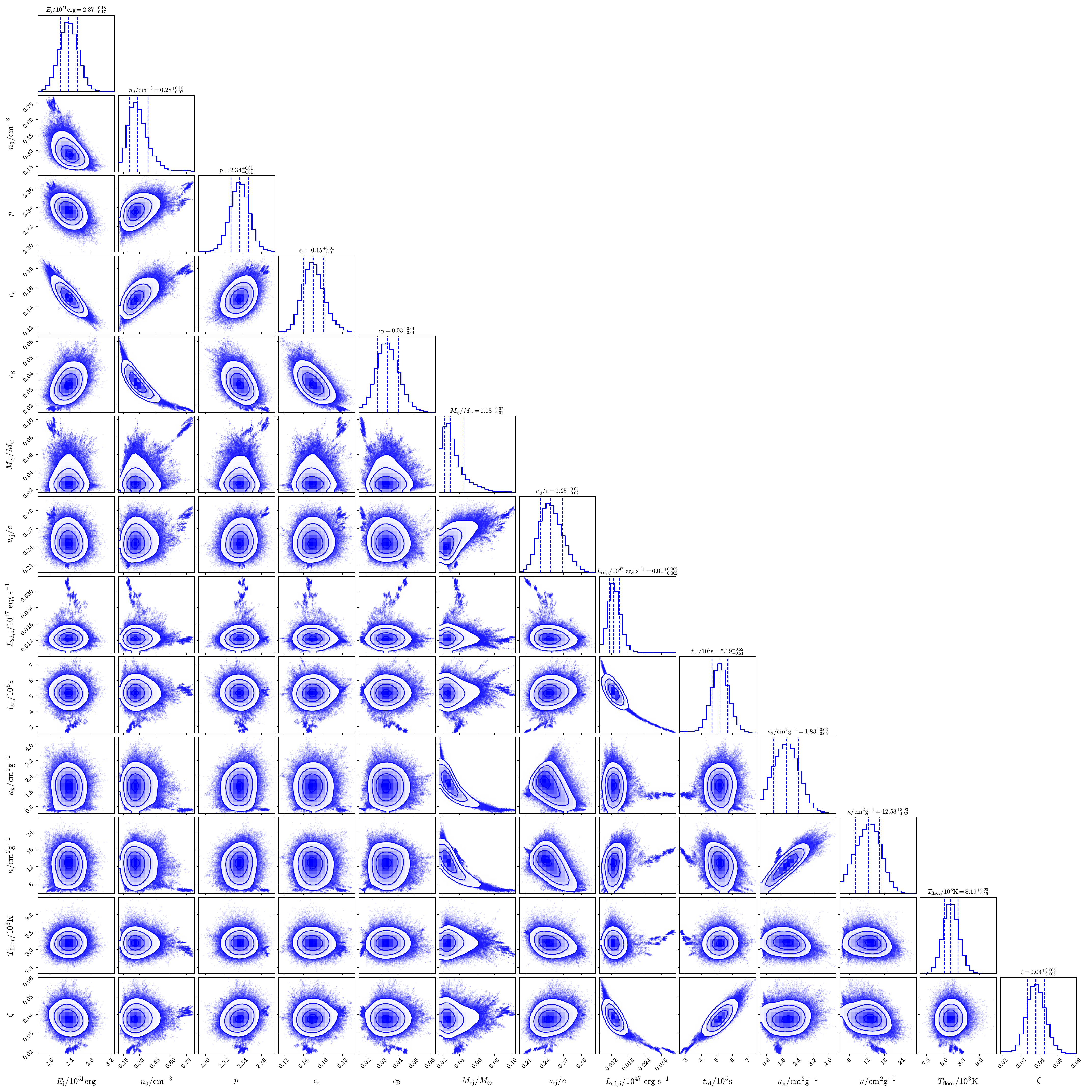}
    \caption{Posteriors of the model parameters for the fitting of EP241021a.}
    \label{fig:corner}
\end{figure*}

\begin{figure*}
    \centering
    \includegraphics[width=0.8\linewidth]{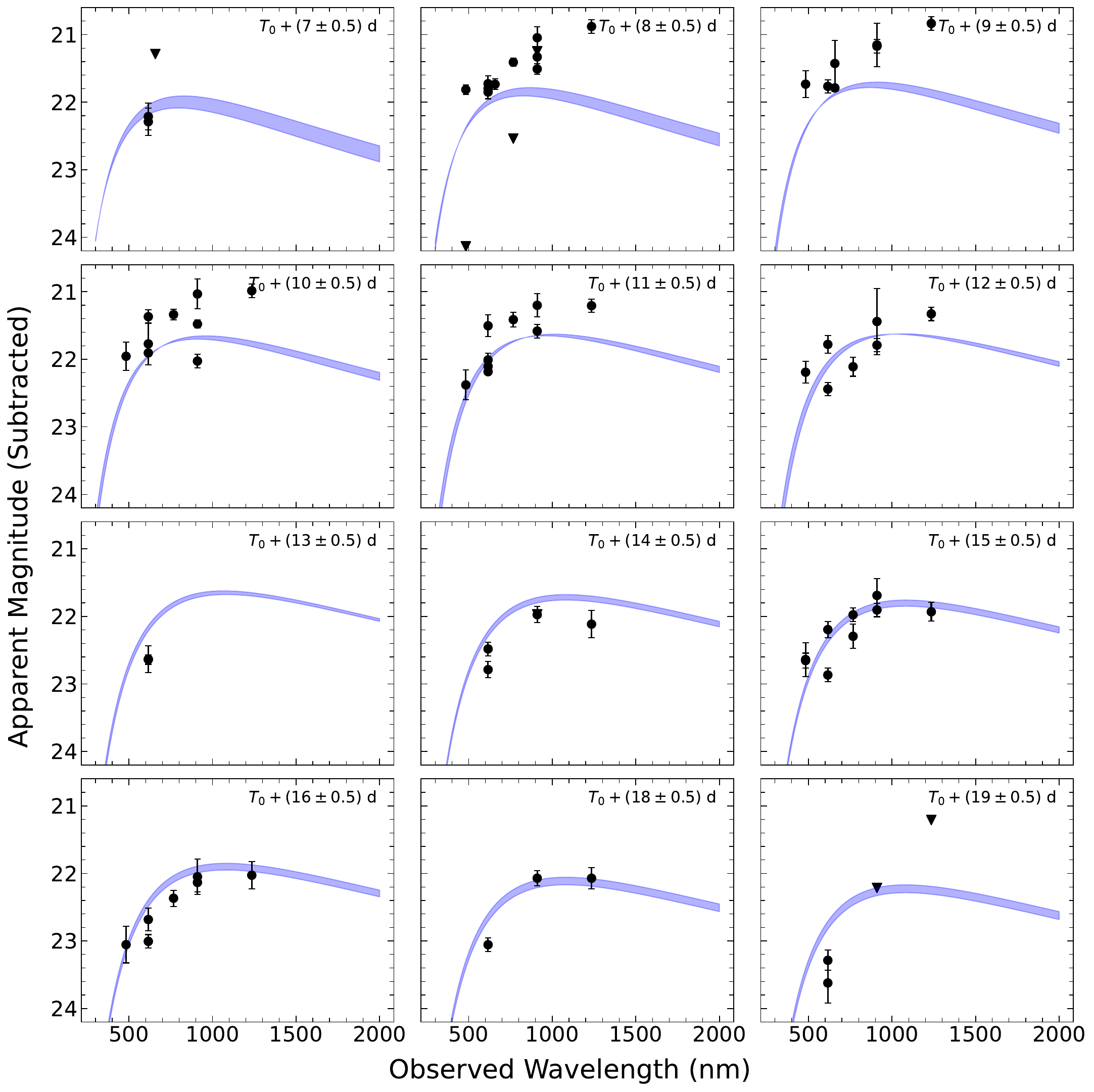}
    \caption{The optical-infrared spectra of EP241021a at different epochs, in comparison with the model-predicted black body spectra. Here, the jet emission component has been subtracted.}
    \label{fig:Optspec}
\end{figure*}

\begin{figure*}
    \centering
    \includegraphics[width=1\linewidth]{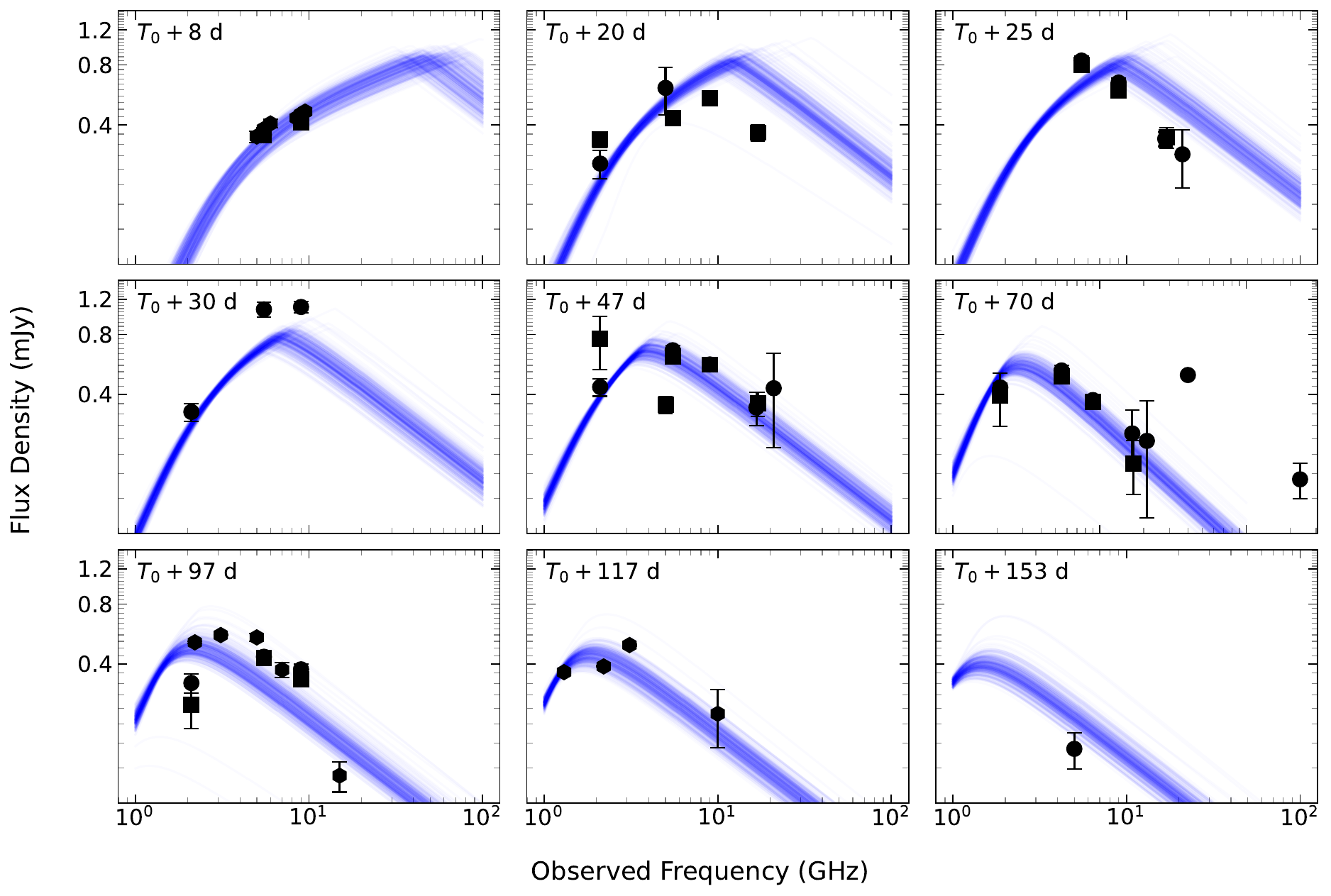}
    \caption{The radio spectra of EP241021a at different epochs. }
    \label{fig:Rspec2}
\end{figure*}

We perform the light curve fittings for EP241021a with the Markov Chain Monte Carlo method of 50000 iterations, by using the \texttt{emcee} package \citep{Foremany2013}. There are five free parameters for the jet emission and eight free parameters relevant to the emission powered by the magnetar. The log likelihood function is $\mathrm{ln}~\mathcal{L}=-0.5(\Sigma_{i=1}^{n}(O_{i}-M_{i})^2/\sigma_{i}^2) $, where $O_{i}$, $M_{i}$ and $\sigma_{i}$ are the $i-\mathrm{th}$ observed flux densities or apparent magnitudes, the model outputs, and the observational uncertainties, respectively. The fitting results are shown by the lines in Figures~\ref{fig:xlc}, \ref{fig:optlc} and \ref{fig:rlc} for the X-ray, optical, and radio light curves, respectively. The priors and posteriors with $16\%$ and $84\%$ quantiles are summarized in Table~\ref{para_tab}. The corner plot of the posterior probability distributions is presented in Figure~\ref{fig:corner}. Furthermore, in Figures~\ref{fig:Optspec}, we further show the general consistency between the model and the optical data in the form of a group of evolving spectra. This indicates that the black body assumption can indeed work for describing the optical bump emission, although its uniqueness for individual spectra should not be overstated. In addition, one may note that, around the peak time $T_0+10$ days, a relatively prominent excess could exist on the red side of the spectra, which can also be found in the light curve fittings. This reveals the complexity of the spectral characteristics of this source. Finally, comparisons between the observational radio spectra and the synchrotron spectra of the jet external shock emission are presented in Figure \ref{fig:Rspec2}.

Generally, our model provides a reasonable explanation for the multi-wavelength observations of EP241021a, including all of the X-ray, optical, and radio emission. It should be noted that we focus on the global evolutionary trends of the multi-wavelength emission, rather than the shorter-timescale fluctuations in light curves. These fluctuations, potentially induced by the inhomogeneous structure of ejecta and internal local activities, are beyond the scope of our simplified model. In any case, the derived parameters for the ejecta agree well with our analyses proposed in Section~\ref{model}. 
The mass, speed, and opacity of the ejecta all direct it towards the origin of a catastrophic compact star collapse/merger. 
The small value of $\zeta$, which is probably smaller than $f_{\rm j}$, further indicates that the energy release from the magnetar could be somewhat anti-beamed in the polar direction but does not deviate from isotropic very much. Even more, the rough equivalence between the isotropic kinetic energies of the jet and the explosion ejecta further strengthens the fact that the magnetar released energy roughly isotropically. This may be a reason why no GRB emission can be detected from the EP241021a event.

\section{Conclusion and Discussion}\label{}
Since the discovery of EP240414a, it is gradually becoming apparent that the launch of EP has brought us to a new era of research of extragalactic optical transients preceded by an FXRT. In view of the relatively high redshift of these transients, their luminosities are inclined to be much higher than those of normal SN-like transients and, thus, probably require an extra energy supply potentially from a central engine. The prompt X-ray emission and the engine-powered optical emission indicate a potential link of these phenomena to GRBs, but simultaneously have significant differences. 

The newly discovered EP241021a was featured by a multi-wavelength rebrightening around ten days after the WXT trigger. The appearance of the X-ray emission indicates that this rebrightening cannot be contributed by an interaction-powered SN\footnote{If only the optical emission was detected, then in principle one cannot rule out that the significant optical bump is contributed by the cooling of SN shock-heated material, because the non-detection of SN signal could just be due to the large distance of the source.}. Alternatively, the bolometric luminosity and peak timescale of the optical bump make it analogous to FBOTs, if its spectra can indeed be described by a black body, which although still needs to be confirmed. %In any case, the relatively red color of the bump indicates it as a new transient phenomenon. 
The primary common feature between EP241021a and FBOTs as well as GRBs could be the existence of a powerful magnetar engine, which may provide us a clue to approach a unified understanding of various extreme stellar explosion phenomena \citep{Yu2017,Liu2022}. As a specific example, EP241021a could hint that the energy release from the rapidly rotating magnetar could not always be collimated in the polar direction on a degree as high as in the GRB case. The anisotropic degree of the energy release of the central engine could be a crucial factor determining the relativistic degree of the jet. Meanwhile, the anisotropic structure of the explosion ejecta would also play an important role in shaping the details of the spectra and light curves of the optical transients, e.g., leading to an obvious excess in the $J$ band.

The reason why EP241021a did not produce GRB radiation but only generated an FXRT might be that the relativistic degree of the jet was not high enough, or it could be that the internal variability in the jet was not significant enough to cause sufficient internal dissipation. It is actually difficult to determine whether the specific mechanism causing FXRT emission is the same as that of GRB, with only differences in energy bands. If so, the duration of EP241021a may tend to support the origins of NS-WD merger and WD collapse, as a DNS merger hardly maintains such a long accretion duration. Furthermore, on the one hand, in contrast to the WD collapse, the NS-WD merger could have an advantage in forming a millisecond magnetar and driving a moderate relativistic jet. On the other hand, if the ambiguous second bump appearing at around $T_0+50$ days is real, as argued by \citep{Shu2025}, then the accretion-induced collapse (AIC) of a WD would become most favored. In this scenario, as proposed by \citet{Zhu2024}, the millisecond magnetar could evaporate the residual companion star quasi-periodically due to their orbital motion. As a result, this evaporated material would provide delayed energy injection into the AIC ejecta, leading to post-peak bumps in the light curve of the ejecta emission. In any case, the advantage of the DNS is that it is most likely to eject sufficient neutron-rich material to explain the high opacity required by the observation. If the prompt X-ray emission is not contributed by the jet, then this possibility cannot be ruled out. If so, EP241021a might be regarded as the first luminous mergernova candidate that was suggested by \citep{Yu2013}.

As proposed by \cite{Zhang2013}, FXRT emission could in principle directly come from the relativistic wind of a magnetar. This has previously been used to explain some observed X-ray orphan transients such as CDF-S XT2 \citep{Xue2019} and some prompt X-ray emission during GRBs \citep[e.g., GRB 230307A;][]{Sun2025}. If the prompt X-ray emission of EP241021a also has such an origin, then its 100s-duration may correspond to an early spin-down timescale of the magnetar. Here, one may note that the fitting of the optical bump has determined $t_{\rm sd}=6$ days, which seems to contradict the above hypothesis. However, the critical constraint on the magnetar parameters from the optical fitting can actually be interpreted as follows: the magnetar should have a spin-down luminosity of $L_{\rm sd,i}=10^{45}\rm erg~s^{-1}$ around the time $t_{\rm sd}=6$ days. For timescales much shorter than 6 days, the optical data cannot provide a strong constraint on whether the spin-down luminosity is the same as at 6 days or significantly higher. In other words, it can not be ruled out that the magnetar may have experienced continuous deceleration within the first several days rather than maintaining a constant spin as assumed. In the GRB 230307A case, it was found that the prompt X-ray emission and the following magnetar-powered kilonova-like emission can be easily connected by a unified spin-down behavior \citep{Wang2024}. However, for EP241021a, since the luminosity of the optical bump is extremely high, it seems difficult to connect the prompt X-ray emission and the optical bump by a simple magnetic braking process. This may also indicate that EP241021a could have an origin different from GRB 230307A which is likely to be produced by an NS-WD merger. In more refined considerations, we need to examine whether the early energy injection might exceed the initial kinetic energy of the ejecta. If so, we must also account for the dynamical evolution of the ejecta in the model.

In any case, if more elaborate modeling is sought in the future, a more complicated braking mechanism could be taken into account. The dynamical evolution of the magnetar wind and its specific emission processes also need to be described carefully \cite[e.g., ][]{Yu2019,Wu2021}, so that a refined X-ray light curve can be obtained. In this case, the abrupt decay of the X-ray emission at $T_0+80$ days, which is indicated by the XMM-Newton upper limits, could be explained by a spectral shift of the magnetar wind emission from the X-ray to lower-energy bands. Such a softening of the wind emission could be a natural result of the decline of the spin-down luminosity and the expansion of the wind nebula. 
%\edit2{due to the expansion of the MWN. This process may also help to explain the late-time excess observed in the optical band relative to our current model predictions.}
Finally, a detailed description of the ejecta structure and frequency-dependent opacity would also help to understand the detailed light-curve and spectral characteristics.

\end{CJK*}
\section*{acknowledgments}
This work is supported by the National SKA Program of China (2020SKA0120300), the National Key R\&D Program of China (2021YFA0718500), the National Natural Science Foundation of China (grant Nos 12393811 and 12303047), Natural Science Foundation of Hubei Province (2023AFB321), and the China Manned Spaced Project (CMS-CSST-2021-A12).

\bibliography{main}{}
\bibliographystyle{aasjournal}

\end{document}